\documentstyle[prl,multicol,aps,psfig]{revtex} 
\begin{document}
%
\draft
\title{Critical Packing Fraction  \\
at the Phase Separation Transition in Hard-Core Mixtures}
\author{Arnaud Buhot
\footnote{buhot@drfmc.ceng.cea.fr}}
\address{CNRS-Laboratoire de Physique Statistique,
Ecole Normale Sup\'{e}rieure,\\
24, rue Lhomond, 75231 Paris Cedex 05, France
\footnote{also at DRFMC/SPSMS CEA Grenoble, 17 av. des Martyrs, 
38054 Grenoble cedex 09, France}}
\date{Received \today}
\maketitle
\begin{abstract}
In this paper, I relate the phase separation transition in binary 
hard-core mixtures in the limit of small size ratio 
to a bond percolation transition.
It allows to estimate the critical packing fraction at the transition 
as a function of the size ratio and the composition of the mixture for 
different shapes of objects. 
The theoretical predictions are in excellent quantitative agreement with 
numerical simulations of binary parallel hard squares mixtures. 
\end{abstract}

\pacs{PACS numbers:  64.75.+g 61.20.Gy}

\begin{multicols}{2}

\narrowtext

Binary mixtures of impenetrable objects pose an easily formulated problem
which is of crucial importance to understand entropic effects in 
colloidal systems.
In this systems of ``large'' and ``small'' objects, an attractive depletion 
force between the large particles results from the presence of the small 
ones~\cite{Asakura}. 
The main interest in such mixtures is the existence of a phase separation 
induced by this effective attraction of large objects.  

Recent Monte-Carlo simulations~\cite{Buhot,Buhot2}
have shown unequivocally that the phase separation 
takes place for spheres, parallel 
cubes and parallel squares for sufficiently dissimilar objects.
Experiments~\cite{Poon} also predict phase separation for mixtures
of hard spheres.
Now that estimates for the critical packing fraction 
at the transition have become available, 
theoretical work has become all the more interesting.
Closures of the Ornstein-Zernicke integral equations have been studied
extensively, and numerical calculations with the Rogers-Young approximation 
predict a demixing transition for sufficiently small size ratios~\cite{Biben}. 
The main drawback of this approach is that the closures
are essentially uncontrolled. 
Other approaches have been used~\cite{Cuesta,Coussaert}
which lead to an instability between
a homogeneous fluid phase and a demixed fluid-fluid phase.
However, as shown by 
numerical simulations~\cite{Buhot,Buhot2} and experiments~\cite{Poon}, 
for small size ratio, the transition is between a fluid phase and a 
fluid-solid phase where 
the solid phase is constituted by closely packed large objects 
surrounded by a fluid of small ones. 
The actual nature of this ``solid'' phase is yet hotly 
debated~\cite{Coussaert,Cuesta2,Dijkstra2}. The aim of this paper is not
to give a better understanding of this phase but to determine the
critical packing fraction at the transition. 

In this article, I propose an expansion for the large-large
pair distribution function in the fluid phase which is expected 
to be convergent in the limit of small size ratio.
If two large particles are sufficiently close so that no
room is left to insert a small
particles between them, they are called {\it bonded}.
This expansion, a virial development, 
allows to calculate the number of {\it bonds} between 
large-large objects, which is an increasing function of the packing fraction. 
Then, I relate the phase separation transition to a bond percolation 
transition and the solid phase to the percolating aggregate.  
Even if this heuristical mapping does not give a better understanding of the
solid phase, it allows to compare the theoretical predictions of the critical 
packing fraction and its dependence on the composition with 
numerical simulations~\cite{Buhot,Buhot2} 
and experiments~\cite{Poon} for objects of different shapes.

In the following, I consider large and small spheres (or disks) of diameters
$\sigma_l$ and $\sigma_s$ and parallel cubes (or squares) of sides
$\sigma_l$ and $\sigma_s$.
The total packing fraction is $\eta = \eta_s + \eta_l$ with $\eta_a = N_a v_a
/ V$, $a = s,l$ where $V$ is the box volume, $N_a$ the number of particles and
$v_a$ the volume of a particle. The size ratio and the composition are
defined as $R =  \sigma_s / \sigma_l$ and $x = \eta_l / \eta$ respectively.
For simplicity, distances are considered in units of the large particle size
($\sigma_l = 1$ so that $\sigma_s = R$).
 
The pair distribution functions are the relevant functions to 
characterize the homogeneous fluid phase. 
The Mayer functions $f_{ab}$ allow to obtain density expansions of the 
large-large pair distribution function $g_{ll}$~\cite{Dijkstra,Hansen}. 
For cubes ($d = 3$) and squares ($d = 2$): 

\begin{equation}
f_{ab}({\bf r}) = - \prod_{i=1}^{d} \Theta ( \sigma_a + \sigma_b - 
2 |r_i|)
\end{equation}

\noindent where $r_i$ stands for the difference of the $i^{th}$ component of 
the positions of particles $a$ and $b$ and $\Theta$ is the Heaviside
step function. For spheres and disks:

\begin{equation} 
f_{ab}({\bf r}) = - \Theta ( \sigma_a + \sigma_b - 2 r)
\end{equation} 

\noindent where $r$ is the distance between the particles $a$ and $b$.
The functions $f_{ab}$ are simply step functions with $f_{ab} = -1$ 
if the two particles $a$ and $b$ overlap and $f_{ab} = 0$ if not.

I propose the following exponential expansion of the pair 
distribution function which is a direct generalization of the one-component
expansion (see Chap.5 in~\cite{Hansen}):

\begin{center}
\unitlength=.2mm
\begin{picture}(360,160)
\put(120,140){\makebox(0,0){$g_{ll}({\bf r}) = \exp \left[ \ 
\eta_{s} R^{-d} a^{-1} \right.$}}
\put(235,150){\circle*{3.}}
\put(225,130){\circle{7.}}
\put(245,130){\circle{7.}}
\put(225,130){\line(1,2){10}}
\put(245,130){\line(-1,2){10}}
\put(280,140){\makebox(0,0){$+ \ \eta_{l} a^{-1}$}}
\put(325,150){\circle*{7.}}
\put(315,130){\circle{7.}}
\put(335,130){\circle{7.}}
\put(315,130){\line(1,2){10}}
\put(335,130){\line(-1,2){10}}
\put(60,100){\makebox(0,0){$+ \frac{1}{2} \eta_{s}^{2} R^{-2d} a^{-2} ( 2$}}
\put(135,110){\circle*{3.}}
\put(155,110){\circle*{3.}}
\put(135,90){\line(0,1){20}}
\put(155,90){\line(0,1){20}}
\put(135,110){\line(1,0){20}}
\put(135,90){\circle{7.}}
\put(155,90){\circle{7.}}
\put(185,100){\makebox(0,0){$+ \  4$}}
\put(210,110){\circle*{3.}}
\put(230,110){\circle*{3.}}
\put(210,90){\line(0,1){20}}
\put(230,90){\line(0,1){20}}
\put(210,110){\line(1,0){20}}
\put(230,90){\line(-1,1){20}}
\put(210,90){\circle{7.}}
\put(230,90){\circle{7.}}
\put(255,100){\makebox(0,0){$+$}}
\put(270,110){\circle*{3.}}
\put(290,110){\circle*{3.}}
\put(270,90){\line(0,1){20}}
\put(290,90){\line(0,1){20}}
\put(270,90){\line(1,1){20}}
\put(290,90){\line(-1,1){20}}
\put(270,110){\line(1,0){20}}
\put(290,90){\circle{7.}}
\put(270,90){\circle{7.}}
\put(305,100){\makebox(0,0){$)$}}
\put(50,20){\makebox(0,0){$+ \frac{1}{2} \eta_{l}^{2} a^{-2} ( 2$}}
\put(105,30){\circle*{7.}}
\put(125,30){\circle*{7.}}
\put(105,10){\line(0,1){20}}
\put(125,10){\line(0,1){20}}
\put(105,30){\line(1,0){20}}
\put(105,10){\circle{7.}}
\put(125,10){\circle{7.}}
\put(155,20){\makebox(0,0){$+ \  4$}}
\put(180,30){\circle*{7.}}
\put(200,30){\circle*{7.}}
\put(180,10){\line(0,1){20}}
\put(200,10){\line(0,1){20}}
\put(180,30){\line(1,0){20}}
\put(200,10){\line(-1,1){20}}
\put(180,10){\circle{7.}}
\put(200,10){\circle{7.}}
\put(225,20){\makebox(0,0){$+$}}
\put(240,30){\circle*{7.}}
\put(260,30){\circle*{7.}}
\put(240,10){\line(0,1){20}}
\put(260,10){\line(0,1){20}}
\put(240,10){\line(1,1){20}}
\put(260,10){\line(-1,1){20}}
\put(240,30){\line(1,0){20}}
\put(260,10){\circle{7.}}
\put(240,10){\circle{7.}}
\put(310,20){\makebox(0,0){$ \left. ) \ + \ \cdots \ \right]$}}
\put(50,60){\makebox(0,0){$+ \frac{1}{2} \eta_{l} \eta_{s} R^{-d} a^{-2} ( 4$}}
\put(130,70){\circle*{3.}}
\put(150,70){\circle*{7.}}
\put(130,50){\line(0,1){20}}
\put(150,50){\line(0,1){20}}
\put(130,70){\line(1,0){20}}
\put(130,50){\circle{7.}}
\put(150,50){\circle{7.}}
\put(175,60){\makebox(0,0){$+ \  4$}}
\put(205,70){\circle*{3.}}
\put(225,70){\circle*{7.}}
\put(205,50){\line(0,1){20}}
\put(225,50){\line(0,1){20}}
\put(205,70){\line(1,0){20}}
\put(225,50){\line(-1,1){20}}
\put(205,50){\circle{7.}}
\put(225,50){\circle{7.}}
\put(250,60){\makebox(0,0){$+ \  4$}}
\put(270,70){\circle*{7.}}
\put(290,70){\circle*{3.}}
\put(270,50){\line(0,1){20}}
\put(290,50){\line(0,1){20}}
\put(270,70){\line(1,0){20}}
\put(290,50){\line(-1,1){20}}
\put(270,50){\circle{7.}}
\put(290,50){\circle{7.}}
\put(315,60){\makebox(0,0){$+ \  2$}}
\put(340,70){\circle*{7.}}
\put(360,70){\circle*{3.}}
\put(340,50){\line(0,1){20}}
\put(360,50){\line(0,1){20}}
\put(340,50){\line(1,1){20}}
\put(360,50){\line(-1,1){20}}
\put(340,70){\line(1,0){20}}
\put(340,50){\circle{7.}}
\put(360,50){\circle{7.}}
\put(375,60){\makebox(0,0){$)$}}
\end{picture}
\end{center}

\noindent where $d$ is the dimension of the space and
$a$ is a constant that depends on the shape of the objects
($a = \pi/6$ for spheres, $a = \pi/4$ for disks and $a = 1$ for cubes 
and squares).
Each diagram is an integral over positions of small and
large particles represented respectively by small and large black dots.
Each link corresponds to a Mayer function and the two large open circles
to two large particles with position difference ${\bf r}$.
For example, the first diagram corresponds to:
 
\begin{center}
\unitlength=.2mm
\begin{picture}(300,40)
\put(20,30){\circle*{3.}}
\put(10,10){\circle{7.}}
\put(30,10){\circle{7.}}
\put(10,10){\line(1,2){10}}
\put(30,10){\line(-1,2){10}}
\put(160,20){\makebox(0,0){$  = \int f_{sl}({\bf r}') f_{sl}({\bf r} -
{\bf r}') d{\bf r}' $}}
\end{picture}
\end{center}
 
\noindent and is a simple function of the position ${\bf r}$ and the size
ratio $R$.
The diagrams in the expansion are all distinct connected diagrams consisting
of two non-adjacent open circles, at least one black dot and such that the
two open circles do not form an articulation~\cite{Hansen}. A diagram with
articulations may be reduced to a product of simpler ones (see below). 

The above expansion, a density development, has negative powers
of $R$ when expressed with the pertinent parameters:
$R$, $x$ and $\eta$. 
However, three different effects may compensate the negative powers: 
first of all, a link between two small particles scales as
$R^d$ since the corresponding Mayer function restricts the integration to a 
small particle volume; second, 
diagrams of the same order but with odd and even numbers of links 
may compensate each other because the Mayer functions are negative;
and last one, the critical packing fraction  at the transition
scales as a positive power of $R$.
Thus, we may expect that the expansion proposed in this article
is convergent when written in powers of the size ratio at the phase 
separation transition. 
In that case, only the first few diagrams have to be considered
in the limit of small $R$.

In the article of Dijkstra et al.~\cite{Dijkstra}, another 
expansion up to the second order was presented for $g_{ll}$.
This expansion is not convergent in the limit of 
small $R$ due to the presence of diagrams which form an 
articulation, like the one with two small particles 
without link between them. In fact, this diagram is the square of a 
simpler one: 

\begin{center}
\unitlength=.2mm
\begin{picture}(150,40)
\put(10,30){\circle*{3.}}
\put(30,30){\circle*{3.}}
\put(10,10){\circle{7.}}
\put(30,10){\circle{7.}}
\put(10,10){\line(0,1){20}}
\put(10,10){\line(1,1){20}}
\put(30,10){\line(0,1){20}}
\put(30,10){\line(-1,1){20}}
\put(50,20){\makebox(0,0){$ = $}}
\put(70,20){\makebox(0,0){$($}}
\put(100,30){\circle*{3.}}
\put(90,10){\circle{7.}}
\put(110,10){\circle{7.}}
\put(90,10){\line(1,2){10}}
\put(110,10){\line(-1,2){10}}
\put(130,20){\makebox(0,0){$)^2 .$}}
\end{picture}
\end{center}

\noindent It scales with a smaller power of $R$ than the others
of the same order and corresponds to the most divergent part
in the second order expansion (see Eq.(5) in~\cite{Dijkstra}). 

\begin{figure}
\centerline{\psfig{figure=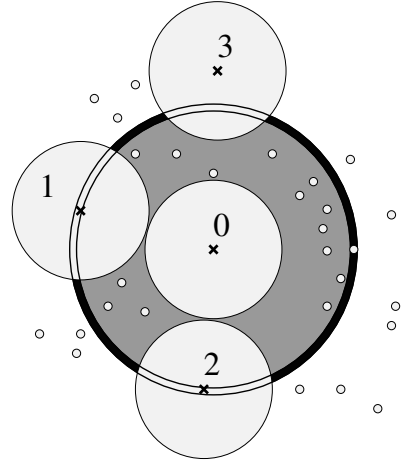,height=7cm}}
\caption{This figure represents the excluded volume for large disks around
the disk $0$ in gray. Large disks are {\it bonded} to the disk $0$
if their centers stay in the ring of width equals to the
small particle diameter around the excluded volume.
In this case, the disks $1$ and $2$ are bonded
to the disk $0$ whereas the disk $3$ is not.
This case corresponds to a size ratio $R = 1/20$.}
\end{figure}

In the binary mixtures, $g_{ll}$ shows a sharp peak at the contact value.
In the limit of small $R$, this peak is as narrow as $R$
and corresponds to large particles with a distance 
$1 \leq r \leq 1+R$ between them.
Two such large particles may be considered as {\it bonded}.
In this case, the mean number $n_b$ of bonds per large particle is 
given by the integration of $g_{ll}$ on the shell of width $R$ around the
excluded volume multiplied by the density of large particles $\rho_l = N_l/V$:

\begin{equation}
n_b = \rho_l \int_{1 \leq r \leq 1 + R}
g_{ll}({\bf r}) d{\bf r}
\end{equation}

\noindent where $r$ is the usual norm of ${\bf r}$ for spheres and
disks, and it is the maximum of the absolute values of the coordinates
of ${\bf r}$ for cubes and squares. In the latter case, it is the relevant
distance since I consider only parallel cubes (or squares).
The example of disks is represented in Fig.1 where the gray region stands for
the excluded volume of large disks around the disk $0$. The ring around the
gray region corresponds to the shell on which the integration is made. 
Centers of disks $1$ and $2$ are in this region 
and those disks are bonded to the disk $0$ but the disk $3$ is not.

From $g_{ll}$, it is clear that
the mean number $n_b$ of bonds per particle is an increasing 
function of the packing fraction. 
Thus, on increasing $\eta$, larger and larger aggregates of 
large particles exist in the system. 
There exists a critical number of bonds per particle $n_c$ for which 
a percolating aggregate appears. 
This corresponds to the phase separation transition and
the percolating aggregate constitutes the solid phase. 
Notice that above this phase separation transition, 
$g_{ll}$ is no longer defined since the translation invariance 
disappears due to the formation of the solid phase.
If we suppose that bonds are randomly distributed, the critical number 
$n_c$ may be related to the bond percolation threshold $p_c$ of the 
lattice corresponding to the close packed configuration. The latter
corresponds to the fraction of bonds necessary to obtain a
percolating aggregate whereas $n_b$ counts the mean number of bonds 
{\it per particle}.
Since the number of bonds per particle in the close packed configuration 
is the coordination number $z$, $n_c = z p_c$.
The fact that positions of objects are not restricted to lattice
sites may slightly modify the critical number $n_c$.
However, $z$ is an obvious upper bound for the critical number $n_c$.

The positions of close packed spheres describe a FCC lattice
of coordination number $z = 12$ and 
the bond percolation threshold of 
this lattice is $p_c = 0.199$~\cite{Stauffer}. 
Thus, the critical number of bonds per particle is $n_{c} = 2.4$.
Disks are close packed in a triangular lattice
($p_c = 0.347$ and $z = 6$)~\cite{Stauffer} which leads to $n_c = 2.1$.
Cubes and squares in a cubic or square lattice ($p_c = 0.249$ or $0.5$ and 
$z = 6$ or $4$ respectively)~\cite{Stauffer} have 
$n_c = 1.5$ or $2$ respectively.
 
In order to obtain the behavior of the critical packing fraction in the 
limit of small $R$, I consider only the first 
diagram of the exponential expansion assuming that the others are negligible. 
In this case, $n_b$ is given by:

\begin{equation}
\label{Eq_nb}
n_b \simeq \frac{\eta_l}{a} \int d{\bf r} \ \exp 
\left( \frac{\eta_s}{a R^{d}} 
\int d{\bf r}' f_{sl}( {\bf r}' ) f_{sl} ( {\bf r} - {\bf r}' ) \right)
\end{equation}

\noindent where the integration over ${\bf r}$ is restricted to $1 \leq r 
\leq 1 + R$.
The critical packing fraction $\eta_c$ is defined by $n_b(\eta_c) = n_c$ 
and depends on the size ratio and the composition. An upper bound 
$\eta_{\rm up}$ for this critical packing fraction is given by the condition 
$n_b(\eta_{\rm up}) = z$.
The estimate for $\eta_c$ is valid only for small $R$, 
for which the peak at the contact value of the pair 
distribution function of large-large particles may be related 
to bonds between large particles and where the diagram considered  
gives the dominant part of $n_b$.

Consider first parallel cubes and squares. The integration of Mayer 
functions $f_{sl}$ is:

\begin{equation}
\int \! d{\bf r}' f_{sl} ({\bf r}')  f_{sl}({\bf r} \! - \! {\bf r}') = 
\prod_{i=1}^{d} (1 \! + \! R \! - \! |r_i|) \Theta (1 \! + \! R \! - \! |r_i|).
\end{equation}

For {\it cubes}, the mean number $n_b$ deduced from this equation and
Eq.(\ref{Eq_nb}) leads to the following critical packing fraction:

\begin{equation}
\eta_c ({\rm Cubes}) \simeq \frac{R^2}{1-x} \ \log \left( \frac{n_c \,
(1-x)}{24 \, x \, R^3} \right).  
\end{equation}

\noindent The upper bound for the critical packing fraction
is given by the same equation replacing $n_c$ by $z$.
Notice that $n_c$ used to calculate $\eta_c$ 
is not essential to obtain the asymptotic behavior 
$-3 R^2 (1-x)^{-1} \log R$ which is also given by 
the upper bound $\eta_{\rm up}$. 
The packing fraction of small cubes $\eta_s = \eta (1-x)$  at the transition
is slightly dependent on composition (through a logarithmic term 
negligible in the limit of $R \rightarrow 0$). 
This result, which is observed for objects of all shapes, is not surprising 
since, in a first approximation, only the small particles are responsible of 
the effective attraction of the large ones. 

The same calculation for the critical packing fraction may be done for 
parallel {\it squares}: 

\begin{equation}
\label{Eq_etac}
\eta_c ({\rm Squares}) \simeq \frac{R}{1-x} \ \log \left( \frac{n_c \, (1-x)}
{8 \, x \, R^2} \right).
\end{equation}

\noindent It is interesting to notice that this is the first 
theoretical prediction of such a phase separation transition 
in binary mixtures of parallel hard squares. 

The behavior of $\eta_c$ as a function of $R$ and $x$ 
has been obtained for the first time in recent simulations
of mixtures of parallel hard squares~\cite{Buhot2}.
Thus, it is possible to compare the theoretical predictions 
with the direct simulations. 
For this case of parallel hard squares, I have been able to push the
expansion considerably farther (to second order) since
the diagrams are quite simple to calculate.
Mayer functions are products of one-dimensional functions and the same 
holds true for the diagrams.
In the following table, theoretical predictions with first order and second
order expansion are compared with simulations.

\begin{center}
\begin{tabular}{|c|c|c|c|c|}
\hline
$R$ & $x$ & \multicolumn{3}{c|}{$\eta_c$} \\
& & $\ 1^{\rm st}$ order \ & $\ 2^{\rm nd}$ order \ & Simulation \\ \hline
$\  1/10 \ $ & $\  0.3 \ $ & $0.68$ & $0.52(0)$ & $0.49 \pm 0.02$ \\ \hline
$ 1/10 $ & $0.5$ & $ \ 0.78 \ $ & $ \ 0.54(4) \ $ & $0.53 \pm 0.02$ \\ \hline
$ 1/10 $ & $0.7$ & $> 1.$ & $0.58(8)$ & $\ 0.60 \pm 0.05 \ $ \\ \hline
$ 1/20 $ & $0.5$ & $0.53$ & $0.42(4)$ & $0.41 \pm 0.02$ \\ \hline
$ 1/50 $ & $0.5$ & $0.29$ & $0.26(4)$ & $0.26 \pm 0.02$ \\ \hline
\end{tabular}
\end{center}

For the first order, a qualitative agreement is obtained 
for $R= 1/10$ and $1/20$, whereas for $R = 1/50$, the agreement is even
quantitative. 
The agreement between numerical simulations and predictions 
with the second order expansion is quite striking, 
and indicates that this exponential expansion succeeds in summing up 
the essential diagrams.
It is interesting to notice that, replacing the packing fraction by its 
critical value (Eq.(\ref{Eq_etac})) in the exponential expansion, 
the second order terms scale at least as $R$ whereas the first order 
term is of order one (neglecting logarithmic terms). 
It would be interesting to prove that higher orders are also negligible.
Moreover, the width of the shell on which the integration is made to
calculate $n_b$ does not seem to be essential as soon as we consider 
a width larger than the peak of the pair distribution function 
($\sim R$) but still of order $R$. 
I have checked for $R = 1/50$ that calculations with shells of width 
$2R$ and $3R$ do not modify significantly $\eta_c$.

Let me now consider the mixture of {\it spheres}. The integration of the Mayer 
functions has already been calculated~\cite{Hansen}:

\begin{equation}
\int d{\bf r}' f_{sl} ({\bf r}')  f_{sl}({\bf r} - {\bf r}') = \frac{\pi}{6} 
(1 \!  + \! R \! - \! r)^2 \, (1 \! + \! R \! + \! r/2)
\end{equation}

\noindent for $1 \leq r \leq 1 + R$. 
Including this expression in Eq.(\ref{Eq_nb}), I deduce the 
critical packing fraction:

\begin{equation}
\label{Spheres}
\eta_c ({\rm Spheres}) \simeq \frac{2 \, R}{3 \, (1-x)} \ \log \left( 
\frac{3 \, (1-x) \, n_c} {8 \, \pi \, x \, R^2} \right)
\end{equation}

\noindent which is valid for $\eta_c \gg 2 \, R / (3 \, (1-x))$. 
This implies that the logarithm of the size ratio is sufficiently large
since $\eta_c$ decreases to zero with $R$ not only linearly but with 
logarithmic corrections.
For $x = 1/2$ and $R = 1/30$, Eq.(\ref{Spheres}) gives  
$\eta_c = 0.24(8)$ whereas numerical simulations~\cite{Buhot} 
have shown that $\eta_c \leq 0.243$. 
Furthermore, both predictions for the critical packing fraction
agree qualitatively with experimental results~\cite{Poon}.
Unfortunately, in order to be able to compare quantitatively 
theoretical with experimental predictions,   
either new experiments with smaller size ratios or theoretical
predictions considering further diagrams in the expansion beyond 
the first one are needed.

An analogous calculation may be done for {\it disks} in two dimensions, 
where the first diagram scales as $R^{3/2}$ in the limit 
$R \rightarrow 0$ for $1 \leq r \leq 1+R$.
This leads to a critical packing fraction: 

\begin{equation}
\label{Eq_disks}
\eta_c({\rm Disks}) \simeq \frac{3 \, \pi}{8 \, \sqrt{2}} 
\frac{\sqrt{R}}{1-x} \log \left(
\frac{\sqrt{2} \, (1-x) \, n_c}{2 \pi \, x \, R^{3/2}} \right).
\end{equation}

\noindent The scaling prediction in $\sqrt{R}$ for $\eta_c$ has been 
already suggested in~\cite{Buhot2},
but there is not yet numerical evidence for the phase separation transition. 
From Eq.(\ref{Eq_disks}), in order to have $\eta_c \sim 0.6$ 
at a composition $x = 0.5$, 
the size ratio to consider is $R \sim 1/600$. 
This seems yet beyond reach since for this size ratio, we would have 
to take into account $360 \  000$ small disks for each 
large one or, in order to clearly see the transition,  
a few dozen of large particles and more than $10^7$ small ones. 
On the other hand, taking into account higher orders in the expansion,
the estimation of the size ratio may be strongly modified.

In conclusion, an exponential expansion of the pair distribution 
function of large-large particles in binary mixtures has been presented. 
A known limitation of such expansion is of course that it is limited
to the homogeneous phase. In this work I have shown that, nevertheless,
this expansion can be used to predict a fluid-solid phase separation by 
computing the quantity which triggers it: a large increase in the
number of large particles in a thin shell around a given large
particle. This was already pointed out in~\cite{Buhot,Buhot2}. 
In the limit of small size ratio, the mean number of bonds between large 
particles at the phase separation transition is related to 
the bond-percolation threshold of the lattice corresponding to the
close packed configuration, and the solid phase corresponds 
to the percolating aggregate of large particles. 
The critical packing fraction is then deduced. 
Its dependence with the composition confirms that the effective attraction 
of large objects is principally due to the depletion of small ones. 
Additional numerical simulations of mixtures of parallel hard squares
reported elsewhere~\cite{Buhot2} confirm
the theoretical predictions.
The theoretical results for mixtures of spheres are in qualitative 
agreement with experiments, whereas the phase separation 
transition for mixtures of hard disks is predicted to occur 
at such small size ratio that the corresponding simulation above the  
transition seems yet out of reach.

It is a pleasure to thank W. Krauth for illuminating discussions and 
M. B. Gordon and W. Krauth for a careful reading of the manuscript.

\end{multicols}
\end{document}